\documentstyle{article}

\begin{document}

\begin{center}
\textbf{COSMOLOGICAL MODELS WITH FRACTIONAL DERIVATIVES AND FRACTIONAL ACTION FUNCTIONAL}\\
\vspace{2,5mm} \emph{\bf{ V.~K.~Shchigolev} \footnote{E-mail:
vkshch@yahoo.com}}
\end{center}
\begin{center}
{\it Department of Theoretical Physics, Faculty of Physics and\\
Engineering, Ulyanovsk State University, Ulyanovsk 432000, Russia}
\end{center}
\bigskip
\begin {abstract}\noindent Cosmological models of a scalar field with
dynamical equations containing fractional derivatives or derived
from the Einstein-Hilbert action of fractional order, are
constructed. A number of exact solutions to those equations of
fractional cosmological  models in both cases is given.

\vspace{2,5mm}


\vspace{2,5mm}

\end {abstract}
\bigskip

\section {Introduction}

\qquad As is observed \cite{C1}, there has been in the last decade active interest to application of
fractional derivatives in various fields of physics,
where they play essential and sometimes leading role in understanding the
complex classical and quantum systems. Surprisingly, the achievements of
the fractional-differential calculus in the theory of gravitation and cosmology are very insufficient
and probably no more then thirty or so publications are devoted to this subject
(see, for example, \cite{C2}-\cite{C7}). At the same time, some essential problems in
modern theoretical cosmology \cite{C8,C9} are known that lead to serious
modification of the possible sources of the accelerated expansion
of the Universe (or the gravitational theory itself).
Under these circumstances, the interest to the
fractional-differential modifications of cosmological models that appeared recently is
fully justified.

In this paper, we provide a review of the main publications
devoted to the fractional differential approach in cosmology
(known to the author). We also offer a novel approach to obtaining
the modified Friedmann equations consistent with main principles
and allowing, in our opinion, to avoid the conflict between the
integer dimension of (pseudo) Riemann space - time of GR and
fractional order of derivatives in the modified equations. As is
mentioned in \cite {C5}, there are two different methods of
approaching what fractional derivative cosmology could be. The
simplest is the Last Step Modification (LSM) method, in which the
Einstein's field equations for a given configuration are replaced
with analogous fractional field equations. In other words, $
\partial _t\to D_t^\alpha$ after the field equations for a
specific geometry have been derived. The fundamentalist
methodology is the First Step Modification (FSM), in which one
starts by constructing fractional derivative geometry. The intensively
developed approach to modification of the main cosmological
equations and non-conservative systems of Lagrangian dynamics on
the basis of a variational principle for the action of a
fractional order (Fractional Action-Like Variational Approach
FALVA) developed in \cite{C2}-\cite{C4} represents one of the
possible version of intermediate modification (Intermediate Step
Approach, ISA) mentioned in \cite{C5}.

Our work is based in its main part on ISA and FALVA, because the
equations of standard cosmology can be obtained from a variational
principle for the Einstein-Hilbert action, in which the variation
is made both over the scale factor and a laps function $N$, or, as
is shown in \cite{C10}, the last can be replaced by a condition of
a time scale invariancy, and in any case it is easy to generalize
masters equations to the case of fractional derivatives. We obtain
the fractional-differential analogue of the Friedmann equations
including a scalar field as a source of gravitation in frameworks
of ISA  and on the basis of FALVA. Until now in cosmology with
fractional derivatives, there exist more questions than answers to
such of them as: How should such models be obtained? How should
the basic equations of these models be written down? What is the
meaning of these solutions for the modified models in aspect of
the modern problems of cosmology? In our work, we try to answer
some of these questions.

\section {Fractional integrals and derivatives}

\qquad There are no less than two dozens of definitions of the
fractional derivative \cite{C1}. In the physical applications of
the fractional differential calculus, more often one deals with
Riemann-Liouville derivative (RLD), Caputo derivative (CD), and
some others.

Such derivatives are defined by means of analytical continuation
of the Cauchy formula for the multiple integral of integer
order as a single integral with a power-law core into the field of
real order $\mu>0$:

\begin{equation}
\label {1}
{}_c I^{\mu}_x f(x)=\frac{1}{\Gamma(\mu)}
\int\limits_{c}^{x} f(t)(x-t)^{\mu-1} dt.
\end{equation}
The Riemann-Liouville derivative of fractional order $\alpha\ge
0 $ of function $f(x)$ is defined as the integer order derivative
of the fractional-order integral (\ref {1}):
\begin{equation}
\label {2}
D^{\alpha}_x f(x)\equiv D^n_x {}_c I^{n-\alpha}_x
f(x)=\frac{1}{\Gamma(n-\alpha)}\frac
{d^n}{dx^n}\int\limits_{c}^{x} \frac{f(t)}{(x-t)^{\alpha-n+1}}dt
\end{equation}
where $D^n_x \equiv d^n/d x^n$,~~$n=[\alpha]+1$. This definition
corresponds to the so-called left derivative, frequently denoted as
${}_c D^{\alpha}_x f(x)$. For the limit $\alpha=1$, this
definition gives $df(x)/dx$. For example, the left RLD of $x^k$
for $\alpha\le 1, c=0$ equals:
\begin{equation}
\label {3}
D^\alpha_x
x^k=\frac{\Gamma(k+1)}{\Gamma(k+1-\alpha)}x^{k-\alpha}.
\end{equation}
For $\alpha=1$, one has the usual result: $D^1_x x^k=k x^{k-1}$.
The interesting feature of RLD is that RLD of non-zero constant
$C_0$ does not equal zero, but for $\alpha\le 1$ it equals
$D^\alpha_x C_0=C_0 x^{-\alpha}/\Gamma(1-\alpha)$. The right RLD
is defined similarly to (\ref{2}) on the interval $[c,d]$:
\begin{equation}
\label {05} {}_x D^{\alpha}_d f(x) =
\frac{1}{\Gamma(n-\alpha)}\left(-\frac {d}{dx}\right)^n
\int\limits_{x}^{d} \frac{f(t)}{(t-x)^{\alpha-n+1}}dt
\end{equation}

The Caputo derivative is the integral transform of a regular
derivative, and it is defined by moving  the integer-order
derivative in the Riemann-Liouville definition (\ref {2}) inside
the integral to act on the function $f(t)$:

\begin{equation}
\label {4} {}^{C}D^{\alpha}_x f(x)\equiv {}_c I^{n-\alpha}_x D^n_x
f(x)=\frac{1}{\Gamma(n-\alpha)}\int\limits_{c}^{x}
\frac{\displaystyle\frac {d^n f(t)}{dt^n}}{(x-t)^{\alpha-n+1}}dt.
\end{equation}
So, for $\alpha\le 0$, $n=1$ è $c=0$, we have the following
expression:
$$
{}^{C}D^{\alpha}_x x^k = \frac{\Gamma(k)}{\Gamma(k+1-\alpha)}k
x^{k-\alpha},
$$
which coincides with RLD $D^\alpha_x x^k$. In general, RLD does
not coincide with CD \cite{C1}:
$$
{}^{C}D^{\alpha}_x f(x)=D^{\alpha}_x
f(x)+\sum^n_{j=0}\frac{1}{\Gamma{(1+j-\alpha)}}\frac{f^{(j)}(c+)}{(x-c)^{\alpha-j}}.
$$
RLD and CD are not defined for power function $x^k$ with
$k=-1$ due to divergency of the integrals in the upper limit. The
derivative of $1/x$ can be obtained as the Weyl derivative defined
as
$$
D^{\alpha}f(x)=\frac{(-1)^{n-1}}{\Gamma{(n-\alpha)}}\int\limits_x^{\infty}\frac
{d^n f(t)}{dt^n}(t-x)^{n-\alpha}dt,
$$
which provides $D^{\alpha}x^{-1}=
-x^{-(1+\alpha)}\Gamma{(1+\alpha)}$  for $\alpha\le1 (n=1)$.

The left and the right CD are defined similarly to those for RLD,
i.e., due to replacing the limits $c\to x$ è $x\to d$ in
definition (\ref{4}) consequently.

One needs to be aware that according to the formulas of addition of
orders, the following holds(\cite{C1}, p.161):
$$
D^{\alpha}_xD^{\beta}_x f(x)= D^{\alpha+\beta}_x f(x)-\sum^n_{j=1}
D^{\beta-j}_x
f(c+)\frac{(x-c)^{-\alpha-j}}{\Gamma{(1-\alpha-j)}}~,
$$
that is $D^{\alpha}_xD^{\beta}_x f(x)\ne D^{\alpha+\beta}_x f(x)$,
if only not all derivatives $D^{\beta-j}_x f(c+)$ at the beginning
of the interval are equal to zero. That is why
$D^{\alpha}_xD^{\alpha}_x f(x)\ne D^{2\alpha}_x f(x)$ in the general
case. Generalizing the Laplace operator in the equation for Newtonian
gravitational potential, the author of \cite{C5} wrongly doubles
the order of the repeated fractional derivative. The authors of
\cite{C11} have avoided this mistake, having written down the
Laplacian $\Delta^{\alpha}$ as:

$$
\Delta^{\alpha}u=\frac{1}{r^{2\alpha}}D^{\alpha}_r(r^{2\alpha}D^{\alpha}_r
u)+\frac{\Gamma^2(\alpha +1)}{r^{2\alpha} \sin^{\alpha}\theta}
\frac{\partial}{\partial\theta}(\sin^{\alpha}\theta \frac{\partial
u}{\partial\theta})+\frac{\Gamma^2(\alpha +1)}{r^{2\alpha} \sin^{2
\alpha}\theta} \frac{\partial^2 u}{\partial \phi}.
$$
However their research did not concern cosmological problems.

Let us note one more property of the fractional derivative
expressed in modification of the Leibniz rule (\cite{C1},p.162):
\begin{equation}
\label {01}
D_x^{\alpha}\left[f(x)g(x)\right]=\sum^{\infty}_{k=0}\frac{\alpha+1}{k!\Gamma{(\alpha-k+1)}}
D_x^{\alpha}D_x^{\alpha-k}f(x)D_x^k g(x),
\end{equation}
which becomes the usual rule as $\alpha=n$. It can be represented
as the integral over the order of fractional derivative:
$$
D_x^{\alpha}\left[f(x)g(x)\right]=\int\limits^{\infty}_{-\infty}\frac{\Gamma(\alpha+1)}
{\Gamma(\mu+1)\Gamma(\alpha+1-\mu)}D_x^{\alpha-\mu}f(x)D_x^{\mu}g(x)d{\mu}.
$$
Later these rules of fractional differentiation will give us an
essential modification to the cosmological models with fractional
derivatives.

At last we want to note, that RLD can be expressed with the help of
the integer derivative at the initial point of the interval and CD
(\cite {C1}, p. 163) due to definition (\ref {4}):
$$
D^{\alpha}_x f(x)=\sum^{n-1}_{k=0}
\frac{(x-c)^{k-\alpha}}{\Gamma{(1+k-\alpha)}}f^{(k)}(c+)~+{}^{C}D^{\alpha}_x
f(x)~.
$$

\section {Cosmological models with fractional derivatives}

\qquad Presumably, for the first time such models within LSM were
considered in \cite {C2}. To avoid a conflict between the occurrence
of fractional derivatives in the Friedmann equations and classical
definition of a tensor in the Einstein equation, based on the
integer order derivative in tensor law of transformation, the quoted
author has repeated the well known derivation of the Friedmann
equations for a dust from the classical approach (see \cite {C3},
\cite {C12}) and than has replaced all integer derivatives with its
fractional analog. For the spatially flat Universe, the Friedmann
equations with a cosmological term  are written down in \cite {C2}
as follows:
\begin{equation}
\label {7}
\begin{array}{rcl}
(D^{\alpha}_t a(t))^2=(A_1 (G \rho)^{\alpha}+B(\Lambda
c^2)^{\alpha})a^2,~~\\
{}\\
D^{\alpha}_t(D^{\alpha}_t a(t))=-(A_2 (G \rho)^{\alpha}-B(\Lambda
c^2)^{\alpha})a,
\end{array}
\end{equation}
where $a(t)$  is a scale factor of the Friedmann-Robertson-Walker
(FRW) line element,
\begin{equation}
\label {8} d s^2 = d t^2- a^2 (t)(d r^2+\xi^2 (r)d \Omega ^2),
\end{equation}
where $\xi (r)=\sin r,r,\sinh r$ for the sign of space curvature
$k=+1,0,-1$, consequently. The occurrence of $ \alpha $ - degrees
in the right-hand side of equations (\ref {7}) is caused,
probably, by dimensional reasons. Then the author of the quoted work
obtained the following  solution to this set in a static case
$a=a_0= constant$:
$$
(G\rho)^{\alpha}=\frac{C_1}{t^{2\alpha}},~~(\Lambda
c^2)^{\alpha}=\frac{C_2}{t^{2\alpha}}.
$$
The conclusion is made, that the density of matter and
cosmological constant decrease as $1/t^2 $ if $G $ and speed of
light in vacuum $c $ remains constant, but if the density of
matter and $\Lambda$ remain constant, then the following holds:
$G\sim 1/t^2 $ and $c\sim 1/t $. The solution of equations (\ref
{7}), mentioned as an illustration of the method of \cite {C2}, in
the form $a (t) =a_0 E _ {1, \alpha} (Ct ^ {\alpha}) $, where

$$
E_{\alpha,\beta}(t)=\sum^{\infty}_{k=0}\frac{t^k}{\Gamma(\alpha
k+\beta)}
$$
is the so-called Mittag-Leffler two-parametric function \cite{C1}, is
not confirmed by any calculations.

It is interesting to note that for $k=0 $ and in the absence of matter
and the cosmological constant, that is if $ \rho=0 $ and $ \Lambda =
0 $, equations (\ref {7}) are reduced to
\begin{equation}
\label {9}
D^{\alpha}_t a(t)=0,~~D^{\alpha}_t(D^{\alpha}_t
a(t))=0,
\end{equation}
which for $ \alpha = 1 $ gives the obvious result: $a (t) =a_0
=constant$, and interval (\ref {8}) is reduced to those of the
Minkowski space. If $ \alpha \ne 1 $, then the substitution of
the zero constant (that is the derivative $D ^ {\alpha} _t a (t) $)
from the first equation of (\ref {9}) into the second one results
in identity irrespectively to the definition of fractional
derivative: RLD or CD. The solution of the first equation in (\ref
{9}) for CD, as well as in the case of integer derivative, is
equal to constant, but for RLD its partial solution is zero, and
the general solution depends on time: $a (t) \sim t ^ {\alpha -1}
$ for $ 0 < \alpha \le 1 $ (see \cite {C1}, p. 216). This
circumstance and also the fact, that initial conditions for
equations with CD should be expressed by means of integer
derivative, instead of fractional one, as in the case of RLD,
frequently decline a choice of definition for the benefit of CD.
In general considerations, we shall not specify a type of
fractional derivative so long as it will be possible.

Let us notice that in \cite {C5} the Universe dynamics equations
are written down in other form, namely:
\begin{equation}
\label {10} 3 [ k +(D^{\alpha}_t a)^2]= \kappa\rho a^2, ~~a^3
D^{\alpha}_t p = D^{\alpha}_t [a^3 (\rho + p)],
\end{equation}
where we use $ \rho $ for the energy density and $p$ for the
pressure. The second equation represents the energy-momentum
conservation law ($T^i _ {j; i} =0 $) for the perfect fluid in
space-time (\ref {8}) with the integer-order derivatives replaced
by its fractional analogues. Assuming further $k=0$ and the
power-law dependence of $a, p $ and $ \rho $ on the time: $a=C
t^n, ~p=A t^m, ~ \rho =B t^r $, with the help of formulas (\ref
{3}) and (\ref {10}) we can find:
$$
m=r=-2
\alpha,~~B=\frac{3}{\kappa}\frac{\Gamma(n+1)^2}{\Gamma(n+1-\alpha)^2},
$$
and then from the equation of state $p = (\gamma -1) \rho $ and
the second equation in (\ref {10}), the following equation for
degree $n$ in the law of evolution of the scale factor is
obtained:
$$
\frac {(\gamma-1)}{\gamma}\frac
{\Gamma(1-2\alpha)}{\Gamma(1-3\alpha)}=\frac
{\Gamma(3n-2\alpha+1)}{\Gamma(3n-3\alpha+1)}.
$$
In the same spirit of naive approach, as it was named by the
author of \cite {C5}, the article \cite {C6} is prepared. In it,
the Riemann curvature tensor and, as a consequence, the Einstein
tensor are defined by the unusual Christoffel symbols containing
fractional (of order $ 0 < \alpha \le 1 $) derivatives of metrics
coefficients,
\begin{equation}
\label {11} \Gamma^{\mu}_{\nu \lambda}(\alpha)=\frac {1}{2}g^{\mu
\nu}(\partial^{\alpha}_{\nu}g_{\rho
\lambda}+\partial^{\alpha}_{\lambda}g_{\rho
\nu}+\partial^{\alpha}_{\rho}g_{\nu \lambda}),
\end{equation}
where $ \partial^{\alpha}_{\nu}$ is a fractional derivative
(\ref{2}) or (\ref{4}) with respect to $x^{\nu}$. The quoted
author writes down the Einstein equation,
$$
R_{\mu \nu}(\alpha)- \frac{1}{2} g_{\mu \nu}
R(\alpha)= \frac{8 \pi G}{c^4} T_{\mu \nu}(\alpha),
$$
and the geodetic equation,
$$
 \frac{d^2 x^{\mu}}{d \tau^2}+ \Gamma^{\mu}_{\nu
\lambda}(\alpha)\frac {d x^{\nu}}{d \tau}\frac {d x^{\lambda}}{d
\tau}=0,
$$
but does not give any solutions, and only demonstrates that
linearized equations in the limit $ \alpha \to 1 $ are reduced to
the usual equations with the integer derivatives for the
gravitational waves and the Newtonian potential, that was expected
from the very beginning due to the definition of a fractional
derivative. Let us note, that an attempt to prove such fractional
derivative replacement in Christoffel symbols, such as (\ref
{11}), is undertaken in \cite {C5} on the basis of
fractional-differential geometry, that would be related to FSM
formalism. However, the quoted author also mentioned that it is so far
not clear what such geometry should be. However, in \cite
{C5} an attempt of construction of fractional geometry with the help
of fractional coordinate transformations $d x^i = D^{\alpha}_jx^i
d x^j$ is undertaken  for the flat two-dimensional space.

There were much more advanced and proved results concerning FSM
formalism and stated by S. Vacaru \cite {C13} (see the
bibliography therein). In those works, the results of construction
of the fractional theory of gravitation for the space - time of
fractional (not integer) dimension are obtained. The author sees
one of the simplest motivation for application of fractional
differential calculus in the theory of gravitation in an
opportunity to avoid singularities of the curvature tensor of
physical meaning due to the completely different geometrical and
physical solutions of the fundamental equations. Besides, it is
noted that models of fractional order are more adapted to
the description of processes with memory, branching and hereditarity,
than those of integer order. The result of the application of the
method developed by the author of nonholonomic deformations to
cosmology was the construction of new classes of cosmological
models \cite {C14}. However, the last does not concern to the
models under consideration, and it will not be considered here.

Let us now mention papers \cite {C3, C4}, where the approach to
the dynamical field theories in general and to the theory of
gravitation is developed on the basis of the variational
principle, formulated by the author, for the action of fractional
order (FALVA). In this approach concerning ISA, the integral of
action $S_L [q] $ for the Lagrangian $L (\tau, q (\tau), \dot q
(\tau)) $ is written down as the fractional integral (\ref {1}):
\begin{equation}
\label {14} S_L [q_i]=\frac {1}{\Gamma
(\alpha)}\int\limits_{t_0}^{t} L(\tau,q_i (\tau),\dot
q_i(\tau))(t-\tau)^{\alpha-1} d\tau ,
\end{equation}
being at fixed $t$ the Stieltjes integral with integrating
function ${\displaystyle g_t(\tau)=
\frac{1}{\Gamma(1+\alpha)}[t^{\alpha}-(t-\tau)^{\alpha}]}$, having
the following scale property:

$$
g_{\mu t}(\mu \tau)=\mu^{\alpha}g_t(\tau),~~\mu>0.
$$
Then $q_i (\tau) $ satisfies the fractional (or modified)
Euler-Lagrange equation:
$$
\frac {\partial L}{\partial q_i}-\frac{d}{d\tau}\Bigl(\frac
{\partial L}{\partial \dot
q_i}\Bigr)=\frac{1-\alpha}{t-\tau}\frac{\partial L}{\partial \dot
q_i}\equiv\overline{F}^i,~~i=1,2,...,n;~~ \tau\in (0,t),
$$
where a dot above the appropriate function stands for the first
derivative with respect to time $ \tau $, $\overline{F}^i$ is the
modified decaying force of "friction", that is the general
expression for non-conservative force. In the article \cite {D1},
time $ \tau $ is treated as the intrinsic (proper) time, and $t $
is the observer time. The author of  \cite {C3, C4}, \cite {C15}
states that at $ \tau \to\infty $ we have $ \overline {F} ^i=0 $,
and provides some examples of application FALVA to the Riedmann
geometry and perturbed cosmological models. Every time the quoted
author does not apply FALVA directly to the gravitational action
$S_G [a (t)] $, where $a (t) $ is the scale factor in the FRW
model (\ref {8}), but tries to take into account influence of the
fractional order action (\ref {14}) on the Friedmann equations
through perturbed (and time-dependent) classical gravitational
constant. Considering Lagrangian $L=g _ {ij} (x, \dot x) \dot
x^i\dot x^j $, the author obtains the modified geodesic equation:
\begin{equation}
\label {16}
\ddot x^i+\frac{\alpha-1}{T}\dot x^i+\Gamma^i_{jk}\dot
x^j\dot x^k=0,
\end{equation}
where $\Gamma^i_{jk}$ are the standard Christoffel symbols, and
$T=t-\tau$.  The second term here is interpreted as a dissipative
force, which infinitely increases as $ \tau \to t $ for $ \alpha
\ne 1 $ and under condition of fixing future  time $t$.
Considering equation (\ref {16}) in the Newtonian approximation,
the author established the time variation of Newton's
gravitational constant (i.e., tried to realize the Dirac
hypothesis\cite {C16}) and introduced the perturbation of the
gravitational constant $ \Delta G = {\displaystyle \frac {3
(1-\alpha)} {4\pi \rho T} \frac {\dot a} {a}} $. Only then, the
effective gravitational constant $G _ {eff} =G +\Delta G $ is
substituted to the standard Friedmann equations:

\begin{eqnarray}
\frac{1}{a^2}({\dot a}^2+k)= \frac{8\pi G_{eff}}{3}\rho +
\frac{\Lambda}{3}{,}\label {17}\\
\frac{\ddot a}{a}= -\frac {4\pi G_{eff}}{3}(\rho+ 3
p)+\frac{\Lambda}{3}{,}\label {18}
\end{eqnarray}
where the cosmological constant $\Lambda$ equals zero or $\Lambda=
(\beta/t)(\dot a/a)$, as it is made in \cite {C17}, where $G _
{eff} =G=const $. Then in \cite {C17}, \cite {C18} the solutions
of equations (\ref {17}), (\ref {18}) are investigated for
various equations of state ($p =\gamma \rho $), which could make
sense if these equations or the way one obtains them were enough justified. It would be
natural to expect of the author of FALVA  the direct
applications of FALVA to construction of the modified Friedmann
equations from the fractional functional $S_G [a (t)] $ that
leads to other results, as it will be shown below, without
necessity to use equation (\ref {16}) for the mentioned above
interpretation of the extra terms in modified equation. Moreover, with the
help of canonical parameter $s=g_t (\tau) $, equation (\ref {16})
is reduced to:

\begin{equation}
\label {19} \ddot x^i+\Gamma^i_{jk}\dot x^j\dot x^k=0,
\end{equation}
where the over dots stand for derivatives with respects to $s$.
Actually, the last means that equations (\ref {16}) could be
obtained without applying FALVA but simply by replacement of
the parameter $s=g_t (\tau) $ in equation (\ref {19}). By the way, it
is possible to address precisely the same remark to \cite {C19},
where the modification of cosmology is undertaken on the basis of
the periodic weight function $g_t (\tau) $ in action:

\begin{equation}
\label {20} S_L [q_i]=\int\limits_{t_0}^{t} L(\tau,q_i (\tau),\dot
q_i(\tau))\exp(-\chi \sin (\beta \tau)) d\tau ,
\end{equation}
which is named fractional in \cite {C19}, though such name
could be applied to functional (\ref {14}) only, meaning its
origination from fractional integral (\ref {1}). As it follows
from (\ref{20}), in this case the weight function is defined by $
{\displaystyle \frac {d g_t (\tau)} {d \tau}} = \exp (-\chi \sin
(\beta \tau))$, and replacement $s=g_t (\tau) $ in (\ref {19}) is
immediately resulted in equation obtained in \cite {C19} by
variation of (\ref {20}):

\begin{equation}
\label {21} \ddot x^i-\beta \chi \cos (\beta \tau)\dot
x^i+\Gamma^i_{jk}\dot x^j\dot x^k=0,
\end{equation}
again with treatment of the second term in (\ref {21}) as the
perturbation of gravitational constant $\Delta G = {\displaystyle
\frac {3\chi \beta \cos (\beta \tau) H} {4 \pi \rho}}$, where the
Hubble parameter $H = {\displaystyle \frac {\dot a} {a}}$.
Certainly, it would be possible  to experiment further with
various weight functions to solve those or other problems of the
gravitational theory and cosmology but it is reasonable before
all to return to the framework stated at the beginning of our
paper and to apply FALVA directly to the gravitational field of
the Universe. By the way, it was possible to understand the
purpose that El-Nabulsi declared while formulating FALVA (see,
for example, \cite {C20}, \cite {C21}  just so).

\section {Fractional derivative cosmology of scalar field}

\qquad First we consider the naive (or LSM) approach to the
fractional derivative cosmological models of a scalar field. For
what follows, it is useful to reproduce the derivation of the
gravitation and scalar field equations from the variational
principle for the Einstein-Hilbert action inspired by ADM
formalism in cosmology (see, e.g. , \cite {C22}). The
Einstein-Hilbert action-like functional for FRW model of the
Universe
\begin{equation}
\label {06} d s^2 = N(t)^2 d t^2- a^2 (t)(d r^2+\xi^2 (r)d \Omega
^2),
\end{equation}
where $N$ is a laps function, filled with a real homogeneous
scalar field $\phi (t)$, is as follows:
\begin{equation}
\label {22} S_{EH}= S_{G}+ S_{Sc}=\frac{3}{8\pi G}\int dtN \left(
-\frac{a \dot a^2}{N^2}+ka-\frac{\Lambda a^3}{3}\right) + \int dtN
a^3 \left(\frac{\dot \phi^2}{2N^2}-V(\phi)\right),
\end{equation}
where $V(\phi)$ is a potential of the field. By variation over
$a(t)$, $\phi (t)$ and $N(t)$ (with the subsequent choice of the
gauge $N=1$) in the action (\ref{22}), one obtains the following
standard Friedmann and scalar field equations:
\begin{eqnarray}
\ddot \phi+3\frac{\dot a}{a}\dot \phi+\frac{d V(\phi)}{d\phi}=0,\label{23}\\
2\frac{\ddot a}{a}+\frac{\dot a^2}{a^2}+\frac{k}{a^2} = -8\pi
G\left(\frac{\dot\phi^2}{2}-V(\phi)\right)
+\Lambda,\label{24}\\
\frac{\dot a^2}{a^2}+\frac{k}{a^2}=\frac{ 8\pi
G}{3}\left(\frac{\dot\phi^2}{2}+V(\phi)\right)+\frac{\Lambda}{3}.\label{25}
\end{eqnarray}
On the other hand, equation (\ref {25}) could be derived with
preliminary gauge $N=1 $ in (\ref {06}), that is with proper time
and representation of the FRW space-time interval as (\ref {8}),
proceeding from time invariancy of the action (\ref {22}), as it
was made in \cite {C10}. As a matter of fact, both approaches are
equivalent, because (\ref {25}) was derived from the
Euler-Lagrange equation for $N (t) $, that is $ {\displaystyle
\frac {\partial L} {\partial N} =0} $, which just expresses noted
time scale invariancy. Besides, instead of equation (\ref {24})
one frequently uses the following equation:
\begin{equation}
\label {26} \frac{\ddot a}{a}= -\frac{8\pi
G}{3}\left(\dot\phi^2-V(\phi)\right)+\frac{\Lambda}{3},
\end{equation}
which turns out from equations (\ref {24}), (\ref {25}).

Considering LSM, it is necessary to make substitution of
fractional derivative of the scale factor and the scalar field
instead of integer derivatives in equations (\ref {23}), (\ref
{25}) and (\ref {26}). Therefore the required set of equations
becomes as follows:
\begin{eqnarray}
D^{\alpha}_t(D^{\alpha}_t \phi) +3\left(\frac{D^{\alpha}_t a}{a}\right)
D^{\alpha}_t \phi + \frac{d V(\phi)}{d\phi}=0{~,}\label{27}\\
\left(D^{\alpha}_t a\right)^2+k=\frac{ 8\pi
G}{3}\left(\frac{1}{2}\left(D^{\alpha}_t
\phi\right)^2+V(\phi)\right)a^2+
\frac{\Lambda}{3}a^2{~,}\label{28}\\
D^{\alpha}_t(D^{\alpha}_t a)= -\frac{8\pi
G}{3}\left(\left(D^{\alpha}_t
\phi\right)^2-V(\phi)\right)a+\frac{\Lambda}{3}a{~.}\label{29}
\end{eqnarray}

While among three equations (\ref {23}) - (\ref {25}) of standard
cosmology only two equations are independent, and the third one
can be derived as the differential consequence of two others, due
to the modified Leibniz rule for the fractional derivative (\ref
{01}), all equations of (\ref {27}) - (\ref {29}) are generally
independent. The last means that one has either to solve equations
(\ref {27}) - (\ref {29}) for $a (t), \phi (t) $ and $V (\phi (t))
$ with constant $G $ and $ \Lambda $ or to admit the dependence of
$G $ and/or $\Lambda $ on time when the potential $V (\phi)$  is
given. This circumstance allows us to doubt the acceptability of
models offered earlier in the frameworks of the naive approach,
which are submitted by the modified equations (\ref {7}) and (\ref
{10}).

The solution of nonlinear fractional equations (\ref {27}) - (\ref
{29}) is rather problematic. Even their classical prototype (\ref
{23}) - (\ref {25}) does not always have exact solutions.
Unfortunately, the theory of equations in fractional derivatives
is incomparably more difficult and less advanced in comparison
with the theory of integer-order differential equations (see,
e.g., {\cite {C1}, \cite {C23}}). Nevertheless, we would like to
give an example of an exact solution to (\ref {27}) - (\ref {29})
demonstrating the fact of existence of some solutions. Let us
consider the spatially flat model of the Universe without
cosmological constant ($k = \Lambda = 0 $). We assume the
dependence of the scale factor, scalar field, and potential on
time as follows : $a = a_0 t^n $, $ \phi = \phi_0 t^m, V (\phi
(t)) = V_0 t^r $. Due to the modified rules of differentiation
(\ref {3}) for the given functions, it follows from (\ref {27}) -
(\ref {29}) that:

\begin{eqnarray}
m = n,~~~r = 2(n-\alpha){~,}\label{30}\\
\phi_0^2 = a_0^2 (4 \pi
G)^{-1}\left(1-\frac{\Gamma(n+1-\alpha)^2}{\Gamma(n+1)
\Gamma(n+1-2\alpha)}\right){,}\label{31}\\
V_0 = a_0^2 (8 \pi
G)^{-1}\frac{\Gamma(n+1)^2}{\Gamma(n+1-\alpha)^2}\left(2+
\frac{\Gamma(n+1-\alpha)^2}{\Gamma(n+1)\Gamma(n+1-2\alpha)}\right){~.}\label{32}\\
V(\phi)=V_0 \frac{2(n-\alpha)}{\phi_0
n}\left(\frac{\phi}{\phi_0}\right)^{\displaystyle
\frac{n-2\alpha}{n}}.\label{33}
\end{eqnarray}

The power $n$ in the dependence of scale factor on time is
connected to the order of fractional derivative $ \alpha $ by the
relationship:
\begin{equation}
\label {34} 2(5 n -2 \alpha)\frac{\Gamma(n+1)}{\Gamma(n+1-\alpha)}
= \left(n+\alpha \pm \sqrt{21 n^2-6 n
\alpha+\alpha^2}\right)\frac{\Gamma(n+1-\alpha)}{\Gamma(n+1-2\alpha)},
\end{equation}
in which both parameters appear in the arguments of $ \Gamma of
$-function, that essentially complicates further general analysis
of the model. Nevertheless, as the scalar field in equation (\ref
{31}) is real, the relationship (\ref {34}) is possible if $n\in
\left (\displaystyle\frac {6} {15} \alpha, \alpha\right) $, where
$ 0 < \alpha < 1 $.

Now we consider the modification of the main equations of the
classical model (\ref {23}) - (\ref {25}), which this set undergoes
at the intermediate approach (or ISA). For this purpose, we
replace the derivatives over time in the action (\ref {22}) with
its fractional (of order $ \alpha $) analogous:
\begin{equation}
\label {35} S_{EH}\equiv\int L_{EH} dt=\int dt N
\left[\frac{3}{8\pi G} \left( -\frac{a (D^{\alpha}_t
a)^2}{N^2}+ka-\frac{\Lambda a^3}{3}\right) +  a^3
\left(\frac{(D^{\alpha}_t \phi)^2}{2N^2}-V(\phi)\right)\right].
\end{equation}

Variational problem with fractional derivatives for functions $q_j
(t) $ in the action $S [q_j] (t) = \displaystyle\int\limits _ {c}
^ {d} L (q_j (t), {} _cD _ {t} ^ {\alpha} q_j (t), {} _t D _ {d} ^
{\beta} q_j (t)) d t $ on the interval $[c, d]$ yields  the
modified Euler-Lagrange equations \cite{C1},\cite{C24}:
\begin{equation}
\label {36} \frac{\partial L}{\partial q_j} + {}_t D_{d}^{\alpha}
\left(\frac{\partial L}{\partial ({}_c D_{t}^{\alpha}q_j)}\right)
+ {}_c D_{t}^{\beta} \left(\frac{\partial L}{\partial ({}_t
D_{d}^{\beta}q_j)}\right) = 0.
\end{equation}
In our case $q_j (t) = \phi (t), N (t) $ and $a (t) $. Therefore
for $L _ {EH} $ from (\ref {35}) and (\ref {36}), we arrive at the
following basic equations of the model:
\begin{eqnarray}
{}_t D^{\alpha}(a^3 D^{\alpha}_t \phi) -
a^3 \frac{d V(\phi)}{d\phi}=0{~,}\label{37}\\
\left(D^{\alpha}_t a\right)^2+k=\frac{ 8\pi
G}{3}\left(\frac{1}{2}\left(D^{\alpha}_t
\phi\right)^2+V(\phi)\right)a^2+
\frac{\Lambda}{3}a^2{~,}\label{38}\\
2 {}_t D^{\alpha}(a~D^{\alpha}_t a)+\left(D^{\alpha}_t
a\right)^2-k=8 \pi G \left(\left(D^{\alpha}_t
\phi\right)^2-V(\phi)\right)a^2-\Lambda a^2{~,}\label{39}
\end{eqnarray}
where we denote ${}_0 D^{\alpha}_t \equiv D^{\alpha}_t$ and ${}_t
D^{\alpha}_{\infty} \equiv {}_tD^{\alpha}$. If in the action $S
[q_j] (t) $ the derivatives are CDs, then (\ref {36}) remains
practically without changes, with the only exception of the
derivatives of Lagrangian in the brackets are CDs \cite {C25}.
Hence if the derivatives $D ^ {\alpha} _t a $ and $D ^ {\alpha} _t
\phi $ in (\ref{35}) are determined as CDs they  also should be
understood as CDs in equations (\ref {37}),(\ref {39}) but the
repeated right derivatives (see definition (\ref {05})) in the
first terms of (\ref{37}) and (\ref{39}) do not undergo
redefinition. Let us note also, that in the case of integer order
derivative ($ \alpha = 1 $) due to the obvious expressions $D^1_t
= \displaystyle\frac {d} {dt} $, $ {} _t D^1 = -\displaystyle\frac
{d} {dt} $, the set of the fractional equations (\ref {37} -\ref
{39}) coincides with the classical one (\ref {23} -\ref {25}).

\section{Cosmological models of scalar field with fractional action}

\qquad We consider now the cosmological model of a scalar field,
which follows from the variational principle for the fractional
action (\ref {14}). This approach is in the framework of the
intermediate approach (ISA) and frequently, as it was mentioned
above, is referred to FALVA. Here, it is necessary to pay
attention to the following peculiar properties of computation to
avoid mistakes in obtaining the master equations of the model
under consideration. The first term in the action (\ref {22}) is
obtained as the result of integration by parts of the first term
in $R\sqrt {-g} $, which depends on the second derivative $ \ddot
a $. Actually this integration removes the derivative of the laps
function $ \dot N $ from the action. It is a well known procedure.
But now, if one makes simply fractional, as (\ref {14}),
generalization of the Einstein-Hilbert action on the basis of
expression (\ref {22}), the result will be incorrect, and some
terms will be omitted. Due to the stated above reason, such terms
are absent in all quoted articles by El-Nabulsi, unfortunately
also containing several other inaccuracies. Therefore, we use the
modified Einstein-Hilbert action $S _ {EH} ^ {\alpha} \equiv
\displaystyle\int \limits _ {0} ^ {t} L _ {EH} ^ {\alpha} \, d
\tau $ as the following fractional integral:
\begin{equation}
\label {40} S_{EH}^{\alpha}=\frac {1}{\Gamma
(\alpha)}\int\limits_{0}^{t} N \left [ \frac{3}{8\pi G}
\left(\frac{a^2\ddot a}{N^2}+\frac{a\dot a^2}{N^2}-\frac{a^2\dot a
\dot N}{N^3}+ka-\frac{\Lambda a^3}{3}\right)+
a^3\left(\frac{\epsilon\dot \phi^2}{2N^2}-V(\phi)\right)\right
](t-\tau)^{\alpha-1} d\tau~,
\end{equation}
where all functions in $L _ {EH} ^ {\alpha} $ depend on the
intrinsic time $ \tau $, and $ \epsilon = +1, -1 $ for the usual
and phantom scalar fields respectively. Varying the action (\ref
{40}) over $q_i = \phi, a $ and $N $ with the subsequent choice of
the gauge $N = 1 $, we obtain the following Euler-Poisson
equations,
$$
\frac{\partial L_{EH}^{\alpha}}{\partial q_i}-\frac{d}{d
\tau}\left(\displaystyle\frac{\partial L_{EH}^{\alpha}}{\partial
\dot q_i} \right)+\frac{d^2}{d
\tau^2}\left(\displaystyle\frac{\partial L_{EH}^{\alpha}}{\partial
\ddot q_i} \right)=0~,
$$
for our model \cite{C26}:
\begin{eqnarray}
\ddot \phi+3\left(\frac{\dot a}{a}+\frac{1-\alpha}{3 t}\right)\dot
\phi+\epsilon
\frac{d V(\phi)}{d\phi}=0{~,}\label{41}\\
\frac{\ddot a}{a}+ \frac{1-\alpha}{2t}\left(\frac{\dot
a}{a}\right)+\frac{(1-\alpha)(2-\alpha)}{2t^2}= -\frac{8\pi
G}{3}\left(\epsilon\dot\phi^2-V(\phi)\right)
+\frac{\Lambda}{3}{~,}\label{42}\\
\left(\frac{\dot a}{a}\right)^2+\frac{1-\alpha}{t}\left(\frac{\dot
a}{a}\right)+\frac{k}{a^2}=\frac{ 8\pi
G}{3}\left(\epsilon\frac{\dot\phi^2}{2}+V(\phi)\right)+\frac{\Lambda}{3}{~,}\label{43}
\end{eqnarray}
where and below the dots above functions designate the derivatives
of the appropriate order with respect to time $t$ obtained by
$t-\tau=T\to t $. Precisely the same equations one can obtain from
generalization of the Euler-Lagrange equations in FALVA for the
Lagrangian with higher (second) derivative (see \cite {C27}) or
from the previous integration by parts as mentioned above. In the
last case, one has to take into account the weight function $
(t-\tau) ^ {\alpha-1} $ in the action integral (\ref {40})
requiring the limit $ \lim _ {t\to 0} (a^2\dot a/t) $ to be
finite.

It is easy to verify that the set of equations (\ref {41}) - (\ref
{43}) is represented by three independent equations, against two
independent equations in the classical case (\ref {23}) - (\ref
{25}), that is connected with violation of the energy - momentum
conservation law $T ^ {\mu\nu} _ {~~; \nu} =0 $. Therefore in the
set of equations (\ref {41}) - (\ref {43}) for three unknown
functions ($a (t), \phi (t) $ and $V (t) $), potential is not an
arbitrary function in general. One can rewrite  equations (\ref
{41}) - (\ref {43}) in terms of effective energy density $ \rho
(t) $ and pressure $p (t) $, taking into account the well known
expressions:
\begin{equation}
\label {45} \rho = \epsilon\frac{1}{2}\dot \phi^2 + V(\phi),~~p =
\epsilon\frac{1}{2}\dot \phi^2 - V(\phi)~.
\end{equation}
As a result, it follows from (\ref{41})-(\ref{43}) and (\ref{45})
that:
\begin{eqnarray}
\dot \rho+3\left(\frac{\dot a}{a}+\frac{1-\alpha}{3 t}\right)(\rho+p)=0{~,}\label{46}\\
\frac{\ddot a}{a}+ \frac{1-\alpha}{2t}\left(\frac{\dot
a}{a}\right)+\frac{(1-\alpha)(2-\alpha)}{2t^2}= -\frac{4\pi
G}{3}(\rho+3p)
+\frac{\Lambda}{3}{~,}\label{47}\\
\left(\frac{\dot a}{a}\right)^2+\frac{1-\alpha}{t}\left(\frac{\dot
a}{a}\right)+\frac{k}{a^2}=\frac{ 8\pi
G}{3}\rho+\frac{\Lambda}{3}{~,}\label{48}
\end{eqnarray}

It is easy to integrate equation (\ref {46}) for the perfect fluid
with equation of state $p = \gamma \rho $:
\begin{equation}
\label {49} \rho =
\frac{\rho_0}{a^{\displaystyle3(1+\gamma)}t^{\displaystyle(1+\gamma)(1-\alpha)}}~.
\end{equation}
The interesting fact is that the cosmological $ \Lambda $-term in
the action (\ref {40}) can depend on time, but in equations (\ref
{41}) - (\ref {43}), and also in (\ref {46}) - (\ref {48}), the
only change appears in the dependence $ \Lambda = \Lambda (t) $.
Therefore the expression (\ref {49}) is also valid.

Let us provide an example of an exact solution for the flat model
($k=0$) and for the quasi-vacuum state of matter: $ \gamma = -1 $.
From (\ref {49}) it follows that $ \rho (t) = \rho_0 =constant $,
as well as in the standard model. Then, the remaining equations of
(\ref {46}) - (\ref {48}) for the Hubble parameter and
$\Lambda$-term can be copied as follows:

\begin{eqnarray}
\dot H - \frac{1-\alpha}{2 t}H +\frac{(1-\alpha)(2-\alpha)}{2t^2}= 0{~,}\label{50}\\
H^2+\frac{1-\alpha}{t} H =\frac{ 8\pi
G}{3}\rho_0+\frac{\Lambda}{3}{~.}\label{51}
\end{eqnarray}
From equation (\ref {50}), it is easy to find that the Hubble
parameter varies with time as
\begin{equation}
\label {52} H = \frac{C_{\alpha}}{t}+ H_0
t^{\displaystyle\frac{1-\alpha}{2}}~,
\end{equation}
where
$C_{\alpha}=\displaystyle\frac{(1-\alpha)(2-\alpha)}{(3-\alpha)}$,
$H_0$ is a constant of integration. From (\ref{52}), the following
dependence of the scale factor on time $t$ appears:
\begin{equation}
\label {53} a = a_0 t^{\displaystyle C_{\alpha}}\exp
\left(\frac{3-\alpha}{2}H_0 t^{\displaystyle \frac{3-\alpha}{2}}
\right) ~,
\end{equation}
while the cosmological $\Lambda$-term changes with time as
\begin{equation}
\label {53} \Lambda = 3 H_0^2 t^{\displaystyle 1-\alpha}+3H_0
\frac{(1-\alpha)(7-3\alpha)}{(3-\alpha)}\phantom{.}t^{\displaystyle-\frac{(1+\alpha)}{2}}+
\frac{3(1-\alpha)^2(2-\alpha)(5-2\alpha)}{(3-\alpha)^2}\frac{1}{t^2}-8\pi
G \rho_0.
\end{equation}

It is obvious that at the proceeding to the standard model in the
limit $ \alpha\to1 $, the obtained solutions (\ref {52}) - (\ref
{53}) reduce to the known exponential expansion of the Universe :
$ a = a_0 e ^ {\displaystyle H_0 t}, ~H=H_0, ~ \Lambda=3 H_0^2
-8\pi G \rho_0 $.

We consider now the dynamics of the flat model of the Universe
($k=0 $) filled by a scalar field. It is convenient to rewrite
equations (\ref {41}) - (\ref {43}) in terms of the Hubble
parameter in the following form:
\begin{eqnarray}
\ddot \phi+3\left(H+\frac{1-\alpha}{3 t}\right)\dot \phi+\epsilon
\frac{d V(\phi)}{d\phi}=0{~,}\label{07}\\
\dot H - \frac{1-\alpha}{2 t}H +\frac{(1-\alpha)(2-\alpha)}{2t^2}=
-  4\pi G\epsilon\dot \phi^2{~,}\label{08}\\
H^2+\frac{1-\alpha}{t} H =\frac{ 8\pi
G}{3}\left(\epsilon\frac{\dot
\phi^2}{2}+V(\phi)\right)+\frac{\Lambda}{3}{~.}\label{09}
\end{eqnarray}
It is easy to prove that the given set of the independent
equations can contain some arbitrariness in a choice of unknown
functions, for instance $H (t) $ or $V (t) $, only if the
cosmological $ \Lambda $-term depends on time. However, it is
possible to proceed from some dependence $ \Lambda (t) $. Let us
consider one example of an exact solution, assuming that the
Hubble parameter and the scalar field depend on time as
follows:
\begin{equation}
\label {54} H = \frac{C}{t},~~ \dot \phi = \frac{B}{t},
\end{equation}
where $C, B $ are the constants to be defined. We take into
account definitions (\ref {45}) and the equation of state $p =
\gamma\rho $ for convenience of representation of solutions in
terms of physically clear parameters and fractional order of the
action $ \alpha $. Using substitution (\ref {54}) in (\ref {07}) -
(\ref {09}) and considering (\ref {45}), we obtain the
following result:
\begin{equation}
\label{55}
 a = a_0{\phantom{.}} t^{\displaystyle\frac{\epsilon}{3}
 \left(\frac{1-\gamma}{1+\gamma}\right)+\frac{\alpha}{3}}~,~~\phi =
\frac{1}{\lambda}\ln t + \phi_0~,~~V(\phi) =
\frac{\epsilon}{2\lambda^2}\left(\frac{1-\gamma}{1+\gamma}\right)
e^{\displaystyle-2\lambda (\phi-\phi_0)}~,
\end{equation}
where the following notation is used:
$$
\lambda =\lambda(\alpha,\gamma) =\pm\sqrt{\frac{24\pi
G}{3-\alpha}}\left[\frac{1-\gamma}{1+\gamma}-2\epsilon\left(\frac{3}{3-\alpha}-2\alpha\right)\right]^{-1/2},
$$
$$
C=\displaystyle\frac{\epsilon}{3}\left(\frac{1-\gamma}{1+\gamma}\right)+\frac{\alpha}{3},~~
B=\lambda ^{-1}.
$$
It is easy to see that the baratropic index $ \gamma $ is not
arbitrary but it should satisfy the following inequality:
$$
\frac{1-\gamma}{1+\gamma}
> 2\epsilon\left(\frac{3}{3-\alpha}-2\alpha\right),
$$
and also the inequality which follows from a weak power condition
$ \rho > 0 $: $ \gamma > -1 $. Thus the $ \Lambda$-term depends
on time $t$ as follows: $ \Lambda = \Lambda_0/t^2 $, where
$$
\Lambda_0 = (1-\alpha)\left[ \frac{(1-\gamma)^2}{6 (1+\gamma)^2}
-\frac{\epsilon (1-\gamma)}{6(1+\gamma)}(9-4\alpha)-1 \right].
$$

The above mentioned solutions demonstrate the existence of
exact solutions to the model, which can be of interest in aspect
of modelling of processes occurring in the Universe. Let us
investigate more generally the properties  of the masters equations
for the spatially flat cosmological model (\ref {07}) - (\ref
{09}). After simple manipulation, this set of equations can be
reorganized as follows:
\begin{eqnarray}
\dot H + 3 H^2  - \frac{2(4-\alpha)}{t}H
-\frac{(1-\alpha)(2-\alpha)}{t^2}
= \frac{t \dot \Lambda}{1-\alpha}{~,}\label{56}\\
4\pi G\epsilon \dot \phi^2 =3 H^2-\frac{3(5-\alpha)}{2 t}H
-\frac{3(1-\alpha)(2-\alpha)}{2 t^2}-\frac{t \dot \Lambda}{1-\alpha}{~,}\label{57}\\
 8\pi G\phantom{.} V(t) + \Lambda = \frac{3(7-3\alpha)}{2 t}H +
 \frac{3(1-\alpha)(2-\alpha)}{2 t^2}{~.}\label{58}
\end{eqnarray}
The term $ \displaystyle \frac {t \dot \Lambda} {1-\alpha} $ in
the right-hand side of  equation (\ref {56}) is obtained by
dividing  the both sides of this equality by $(1-\alpha)$.
Therefore equation (\ref {56}) is valid for all $ \alpha \in (0;
1) $ but in the classical limit $ \alpha \to 1 $ it simply means $
\dot \Lambda = 0 $, i.e. $ \Lambda = constant $. The latter is the
only consequence of $ \alpha = 1 $, and among three equations
(\ref {56}) - (\ref {58}) it is necessary to solve only two of
them with $ \dot \Lambda = 0 $, that allows us to define one of
the function among $H (t), \phi (t) $ or $V (\phi)$ arbitrarily.

The situation is essentially different for $ \alpha \ne 1 $. Now
we have three independent equations for four functions: $H (t),
\phi (t), V (\phi) $ and $ \Lambda (t) $, and specification of $
\Lambda (t) $ allows us to find the dependence $H (t)$ from
equation (\ref {56}). After that, it is possible to find $ \phi
(t) $ and $V (t)$ from equations (\ref {57}) and (\ref {58}). It
is important that the evolutionary, i.e. containing derivative
over time, equation (\ref {56}) for the Hubble parameter
experiences the influence of the parameter $ \Lambda (t)$ only.
Thus, the behaviour of the field and its potential becomes
secondary, not determining an expansion dynamics of the Universe
directly, and can be simply found from the equations (\ref {57}),
(\ref {58}). In some sense, the scalar field here plays a role of
a latent parameter. Indeed, the system behaviour is determined by
the field behaviour but the Hubble parameter and the scale factor
are determined not directly by the scalar field but only through
the $\Lambda (t)$-term.  Using some phenomenological expression
for the cosmological parameter, it is possible to find the
dependence of expansion on time, and then the scalar field that
provided it. There is a certain sense in that, as we know
practically nothing about a scalar field that evoked cosmological
inflation and the present accelerated expansion of the Universe.

We consider one example of the mentioned approach, assuming $
\Lambda = constant$. In this case, equation (\ref {56}) contains
the only free parameter: $ \alpha $, and can be easily solved,
that gives the Hubble parameter as:
\begin{equation}
\label{59} H = \frac{1}{6 t}\left (9-2 \alpha
-w_{\alpha}\frac{1-c_0 t^{\displaystyle
w_{\alpha}}}{1+c_0\displaystyle t^{\displaystyle
w_{\alpha}}}\right ),~~~w_{\alpha}=\sqrt{16
\alpha^2-72\alpha+105}~,~~~c_0>0.
\end{equation}
One can notice, that $w _ {\alpha} > 0 $ for $ \alpha\in (0; 1)$.
Integrating equation (\ref {59}), we obtain the following
expression for the scale factor:
\begin{equation}
\label{60}
 a = a_0{\phantom{.}} \left\{t^{\displaystyle(9-2\alpha-w_{\alpha})/2} \displaystyle\left(1+c_0
\displaystyle t^{\displaystyle w_{\alpha}}\right)\right\}^{1/3}.
\end{equation}
Substitution of the Hubble parameter from equation (\ref {59}) in
(\ref {57}), (\ref {58}) gives us expressions for $ \phi (t) $ and
$V (t) $. Due to the definition (\ref {45}) and equations ( \ref
{57}), (\ref {58}) the energy density and the pressure of the
scalar field ($ \Lambda = 0 $) are as follows:
$$
 \rho = \frac{1}{96 \pi G t^2}\left (9-2 \alpha
-w_{\alpha}\frac{1-c_0 t^{\displaystyle
w_{\alpha}}}{1+c_0\displaystyle t^{\displaystyle
w_{\alpha}}}\right )\left (15-8 \alpha -w_{\alpha}\frac{1-c_0
t^{\displaystyle w_{\alpha}}}{1+c_0\displaystyle t^{\displaystyle
w_{\alpha}}}\right ),
$$
$$
 p = \frac{1}{96 \pi G t^2}\left[\left (9-2 \alpha
-w_{\alpha}\frac{1-c_0 t^{\displaystyle
w_{\alpha}}}{1+c_0\displaystyle t^{\displaystyle
w_{\alpha}}}\right )\left (10 \alpha-27 -w_{\alpha}\frac{1-c_0
t^{\displaystyle w_{\alpha}}}{1+c_0\displaystyle t^{\displaystyle
w_{\alpha}}}\right )-36(1-\alpha)(2-\alpha)\right].
$$
From these expressions, it follows that the barotropic index $
\gamma =\rho/p $ is not constant, and the equation of state
changes with time. It is possible to show that at the initial
moment of time, $ \gamma (0) $ strongly depends on the fractional
order $ \alpha $, but at $t\to\infty $ it practically loses this
dependence, slowly approaching $ - 5/7 $ as $ \alpha \to 1 $.

It is interesting that if we assume the cosmological term decays
as $\Lambda = \displaystyle \frac{\beta}{t}H$, where $ \beta $ is
a positive constant, then equation (ref {56}) is reduced into the
following one:
$$
(1-\alpha - \beta)\dot H + 3(1-\alpha) H^2  -
\frac{[2(4-\alpha)(1-\alpha)-\beta]}{t}H
-\frac{(1-\alpha)^2(2-\alpha)}{t^2} = 0~,
$$
differing from the same equation in the case $ \beta=0 $,
considered above, only by the constant multipliers. Therefore its
solution is structurally similar to (\ref {59}), (\ref {60}).

\section{Conclusion}

\qquad In our article, we have given the critical analysis of the
main results and approaches known by now that are directed towards the
application in cosmology of the ideas of fractional
differentiation and integration. At the same time, we offer the
unified consecutive approach to such a problem  on the basis of
the variational principle for the Einstein-Hilbert action which
either includes fractional derivatives of the scale factor and
the scalar field in the Lagrangian, or becomes fractional one. The exact
solutions of the modified Friedmann and scalar field equations
essentially differ from the ones known earlier that could be useful in
aspect of modifications undertaken in modern cosmology in view of
new observational puzzles \cite{C28}. We hope that further
investigation of the obtained equations and its solutions will
allow us at last to answer the question: in what degree the ideas of
the fractional differential calculus are productive in cosmology.

The author is grateful to Prof. V.V. Uchaikin for the offer of the
topic and the valuable discussions.


\begin{thebibliography}{99}

\bibitem{C1}V.V. Uchaikin. Fractional Derivatives Method.-Ulyanovsk,"Artishok", 2008 (in russian).
\bibitem{C2} A.R. El-Nabulsi, Fractional Unstable Euclidean Universe/ EJTP 8 (2005)1-11, EJTP 5 No.17(2008)103- 106.
\bibitem{C3} A.R. El-Nabulsi,Differential Geometry and Modern Cosmology with Fractionaly Differentiated Lagrangian
Function and Fractional Decaying Force Term/ Rom.Journ. Phys.,
Vol.52, Nos. 3-4 (2007) pp.467-481.
\bibitem{C4}A.R. El-Nabulsi, Cosmology with a Fractional Action Priciple/ Rom. Report in Phys., Vol.59, No. 3 (2007)
pp.763-771.
\bibitem{C5}M.D. Roberts, Fractional Derivative Cosmology/ arXiv: 0909.117 [gr-qc].
\bibitem{C6}J.Munkhammar, Riemann-Liouville Fractional Einstein Field Equations/ arXiv: 1003.4981 [physics.gen-ph].
\bibitem{C7}S.I. Vacaru, Fractional Nonholonomic Ricci Flows/arXiv:1004.0625 [math.DG].
\bibitem{C8}V. Sahni, A. Starobinsky, Reconstrucing Dark
Energy/arXiv: 0610026 [astro-ph]
\bibitem{C9} V.N. Lukash, V.A. Rubakov, Dark Energy: Miths and Reality/UFN (2008), v.178, ¹ 3, pp. 301-308 (in russian).
\bibitem{C10}V.V. Kiselev, Vector field as a quintessence partner/arXiv: 0402095 [gr-qc].
\bibitem{C11} M.-F. Li, J.-R. Ren, T. Zhu, Fractional Vector Calculus and Fractional Special
Function/arXiv: 1001.2889v1 [math-ph]
\bibitem{C12} J.W. Norbury, From Newton's Laws to the
Wheeler-DeWitt Equation/European Journal of Physics,(1998) vol. 19,
pg. 143-150.
\bibitem{C13} S.I. Vacaru, Fractional Dynamics from Einstein Gravity, General Solutions,
and Black Holes/arXiv:1004.0628[math-ph].
\bibitem{C14} S.I. Vacaru, New Classes of Off-Diagonal Cosmological Solutions in Einstein Graviti/arXiv:1003.0043[gr-qc].
\bibitem{D1} G.S.F. Frederico, D.F.M. Torres,  Constants of motion for fractional action-like variational problems /
arXiv: 0607472[math.OC].
\bibitem{C15} A.R. El-Nabulsi, Fractional Action-Like Variational Problems/arXiv: 0804.4500 [math-ph].
\bibitem{C16} P.A.M. Dirac, Nature {\bf 61} , 323 (1937).
\bibitem{C17}A.R. El-Nabulsi, Accelerated Expansion of a Closed Universe/ Fizika B (Zagreb), {\bf 16},3, 167-174
(2007).
\bibitem{C18} A.R. El-Nabulsi, Some Fractional Geometrical Aspects of Weak Field Approximation and
Schwarzschild Spacetime
/ Rom.Journ. Phys., Vol.52, Nos. 5-7, 705-715 (2007).
\bibitem{C19} A.R. El-Nabulsi, Oscillating Flat FRW Dark Enegy
Dominated Cosmology from Periodic Functional / Commun. Theor.
Phys. (Beijing, China), Vol. 54, No. 1 (2010) pp. 16-20.
\bibitem{C20} A.R. El-Nabulsi, A fractional action-like
variational  approach of some classical, quantum  and geometrical
dynamics / Int.J. Appl. Math. 17, No 3 (2005), 299-317.
\bibitem{C21} A.R. El-Nabulsi, Increasing Effective Gravitational Constant in Fractional ADD Brane Cosmology /
EJTP {\bf 5}, No.17 (2008), pp. 103-106.
\bibitem{C22} C. Kiefer and B. Sandh\"ofer, Quantum Cosmology/arXiv: 0804.0672 [qr-qc].
\bibitem{C23} I. Podlubny. Fractional Differential Equations. - New York - London: Academic Press, 1999.
\bibitem{C24} D. Baleanu, S.I. Muslih, Lagrangiuan formulation of classical fields within Riemann-Liouville
fractional derivatives/ arXiv: 0510071 [hep-th].
\bibitem{C25} D. Baleanu, Om. P. Agrawal, Fractional Hamilton formalism within Caputo
derivative/ arXiv: 0612025 [math-ph].
\bibitem{C26} V.K. Shchigolev, Fractional-differential cosmological
models// Int. Conference " Modern Problems of Gravitation,
Cosmology and Relativistic Astrophysics" (RUDN-2010), 27 June- 3
July 2010, PFUR, Moscow, Russia, p.75.
\bibitem{C27} G.S.F. Frederico, D.F.M. Torres,  Necessary Optimality Condition for Fractional Action-Like
Problems with Intrinsic and Observer Times/ arXiv: 0712.0152
[math.OC].
\bibitem{C28} S. Pelmutter {\it et al}. [Supernova Cosmology Project Collaboration], Astrophys. J. {\bf 517}, 565
(1999).
\end{thebibliography}
\end{document}